# Post-Synthetic Treatment of Nickel-Iron Layered Double Hydroxides for the Optimum Catalysis of the Oxygen Evolution Reaction

Daire Tyndal[1,2], Sonia Jaskaniec[1,2], Brian Shortall[2,3], Ahin Roy[1,2], Lee Gannon[2,3], Katie O'Neill[1,2], Michelle P. Browne[1,2], João Coelho[1,2,4], Cormac McGuinness[2,3], Georg S. Duesberg[1,2,5], and Valeria Nicolosi[1,2,6]*

[1]School of Chemistry, Trinity College Dublin, Dublin, Ireland

[2]CRANN & AMBER, Trinity College Dublin, Dublin, Ireland

[3]School of Physics, Trinity College Dublin, Dublin, Ireland

[4]CENIMAT|i3N Departamento de Ciência de Materias, Faculdade de Ciências e Tecnologia, Universidade NOVA de Lisboa and CEMOP/UNINOVA, Campus da Caparica, Portugal

[5]Faculty of Electrical Engineering and Information Technology, Universität der Bundeswehr München, Neubiberg, Germany

[6]I-Form research, Trinity College Dublin, Dublin, Ireland

**E-mail**=nicolov@tcd.ie

## Abstract

Nickel-iron layered double hydroxide (NiFe LDH) platelets with high morphological regularity and sub-micrometre lateral dimensions were synthesized using a homogeneous precipitation technique for highly efficient catalysis of the oxygen evolution reaction (OER). Considering edge sites are the point of activity, efforts were made to control platelet size within the synthesized dispersions. The goal is to controllably isolate and characterize size-reduced NiFe LDH particles. Synthetic approaches for size control of NiFe LDH platelets have not been transferable based on published work with other LDH materials and for that reason, we instead use post-synthetic treatment techniques to improve edge-site density. In the end, size reduced NiFe

LDH/single-wall carbon nanotube (SWCNT) composites allowed to further reduce the OER overpotential to 237±7 mV ($<L>$ = 0.16±0.01 $\mu$m, 20 wt% SWCNT), which is one of the best values reported to date. This approach as well improved the long term activity of the catalyst in operating conditions.

# 1 Introduction

One of the most significant concerns in modern society is the imminent energy crisis which will be faced if fossil fuels cannot be successfully replaced by some renewable and clean alternative energy source[1, 2]. At the moment, $H_2$ gas seems to be a viable option, as it combusts cleanly in air and has more chemical energy per unit weight than gasoline and common battery materials[1]. Unfortunately, $H_2$ gas is not readily available in nature. The cheapest and most popular synthetic route for industrial scale hydrogen production to date is steam reforming using natural gas[3]. A cleaner method of production can be achieved via the extraction of $H_2$ from water by electrolysis[1, 2, 4]. Water splitting consists of two half reactions, the hydrogen evolution reaction (HER) and oxygen evolution reaction (OER)[1, 2, 4]. While, HER is a relatively simple process, OER is characterized by sluggish kinetics[2, 4, 5]. Thus, in practice OER is observed at applied potentials higher than its predicted equilibrium potential[4]. This excess overpotential renders the overall water electrolysis an energetically costly process[6, 7]. To this day, water electrolysis accounts for just 4% of the world's hydrogen production because the energy cost remains high, meaning more efficient catalysis of the OER reaction is imperative[1, 2, 4, 8]. Layered Double Hydroxides (LDH) are among the materials of highest interest in research today, owing to their desirable chemical and electrochemical properties, with applications in electrocatalysis,[1, 2, 4, 9–12] as well as energy storage and drug delivery[13]. In particular, nickel-iron (NiFe) LDH is a material with impressive catalytic properties as a result of a high density of active-sites whose activity is enhanced by a charge-transfer activation mechanism between the Ni and Fe metal centres[14]. Moreover, the LDH's relatively open brucite-like structure promotes as well a fast diffusion of reactants and products during catalysis[9, 10]. For comparison purposes, in terms of OER overpotential and stability in operating conditions, NiFe LDH outperforms competitive and more conventional catalyst materials, such as $RuO_2$ and $IrO_2$ as well as Raney Nickel (Alkaline Water Electrolysis), while also being significantly cheaper[1, 2, 11, 15, 16]. Not surprisingly, the development of synthesis routes[17, 18] had lead to new and improved NiFe LDH based catalysts, such as molecule intercalated LDH[18], nanoparticles[2, 19–21] and NiFe LDH/carbon hybrids[19–21], among others[22, 23]. In its pure form NiFe LDH can comfortably produce OER overpotentials below 400 mV (@ 10 mA·cm⁻²) which can be pushed to considerably lower values with various novel systems. For example, Song *et al*[24] achieve 300 mV by exfoliation alone while Gong et al[25] demonstrate 220 mV for NiFe LDH platelets grown directly onto a carbon nanotube (CNT) binder. All of these attempts aim at increasing the number of accessible active sites and matrix integration, hence improving the overall catalytic activity. In theory, reducing the particle size should at some stage result in an optimized OER activity. However, OER activity optimization as a function of the particle size is not commonly reported in literature[26]. The lack of focus in literature is perhaps derivative of the still ambiguous nature of the material's activity. Different theories are being published in attempts to prove the nature of catalytically active sites and the nature of NiFe LDH's OER activity through time[27, 28]. This work aims to present a systematic study as to the extent to which particle size can be used to enhance the catalytic abilities of NiFe LDH and morphologically similar catalysts for oxygen evolution by water electrolysis. Additionally, the compatibility of the size-controlled material of NiFe LDH with composite additives will be investigated for even higher OER activity. Carbon materials such as CNT[20, 21], graphene[29–31] and graphene oxides (GO)[19, 32] are often considered as desirable components in electrocatalyst composite materials due to their high conductivity and impressive mechanical properties. However, it is important to consider that carbon materials may corrode significantly at high potentials and are also known to contain impurities including Ni and Fe which may act as sites for OER. Such considerations are held in regard in this work when testing carbon composite catalysts. In this work, single-wall carbon nanotubes (SWCNT) are investigated due to their strong, flexible, lightweight nature which makes them potentially the most compatible additive for the (sr)-NiFe LDH platelets. Ultimately, identifying an appropriate composite system should produce a somewhat optimized catalyst of the material with competitive OER behavior.

## 2 Results and Discussion

### 2.1 Synthesis and characterization of NiFe LDH platelets

NiFe LDH was synthesised by way of homogeneous coprecipitation of $Ni^{2+}$ and $Fe^{3+}$ and using a triethanolamine (TEA) capping agent. This synthetic route has been shown to yield pure, highly crystalline NiFe LDH hexagonal platelets with planar morphology[17], as shown in Figure 1a. The platelets exhibit sharp edges and relatively homogeneous surface, with some patches indicating the presence of minor amounts of residual TEA, on the surface which have accumulated during preparation[17]. The sharp hexagonal features and regular planar morphology have significance here because they are qualities which have been shown to improve catalytic performance for the OER relative to other morphologies which may result from different synthetic routes[17]. Combined quantitative compositional thermogravimentric analysis (Supplementary Figure 1) and atomic absorption spectroscopy approaches give an estimated molecular formula $Ni_{0.78}Fe_{0.22}(OH_2)(CO_3)0.11$-$0.5H_2O$.

The X-Ray Diffraction (XRD) pattern shown in Figure 1b was indexed as hexagonal symmetry with lattice parameters a = 3.08 Å, c = 23.55 Å. The high intensity peaks in the diffractogram such as (003) represent high-symmetry directions within the crystal and suggest a highly crystalline nature of the sample[24]. The accurate peak matching also suggests that the sample is pure. The lattice parameters are consistent with the 7.8 Å interlayer spacing expected between layers of $CO^{2-}$ intercalated NiFe LDH[33]. Two unmatched peaks (labelled * in Figure 1b) suggest the presence of some minor swollen phase relative to the dominant $CO^{2-}$ intercalated material. It is known that some residual TEA is present within the synthesized material, as evidenced by Jaskaniec et al.[17], characterized with infrared spectroscopy. Given the otherwise high degree of purity, it is reasonable to suggest that TEA intercalation is responsible for the minor phase. Additional atomic-resolution HAADF-STEM imaging (Figure 1 c) indicates a defect-free, crystalline surface, while the hexagonal orientation of the bright SAED pattern (inset) is characteristic of the symmetry within a highly ordered LDH structure[24]. Coupling this with XRD data, the first two families of bright spots can be indexed as (10-10) and (11-10) directions. These results confirm that the used synthesis route maximises the products crystallinity and morphological regularity for this synthetic approach. NiFe LDH platelets produced as such can significantly enhance the OER activity of a nickel foam catalyst when the material is homogeneously deposited on the the surface (Supplementary Figure 2).

## 2.2    Mechanical size-reduction of NiFe LDH platelets

SEM micrographs (Figure 2a) over an extended sample area, reveal that well defined hexagonal platelets have been formed throughout with a high degree of regularity in terms of platelet morphology and size. A particle size study was carried out by measuring individual particle lateral dimensions within a sample set between adjacent corners (Figure 2b and Supplementary Methods). This study revealed that as-produced hexagons had an average lateral length 0.78±0.2 $\mu$m, with the majority of the flakes existing in a relatively narrow size range of 0.4 - 1 $\mu$m. This is consistent with sub-micron NiFe LDH lateral dimensions reported in the literature (smaller than most LDH species) which is known to be one of the materials significant features in terms of electrochemical activity[24]. Ultrasonication of the as-produced platelets using a tip-sonication method can be used to further reduce the average particle size. When high power tip-sonication is applied, the shear stress required for exfoliation can be surpassed and out-of-plane particle breaking dominates. In this way, fragmentation of the LDH platelets can be achieved, resulting in average particle size reduction. Figure 2c and Figure 2d reveal that the original NiFe LDH hexagons can be reduced to platelets of 0.29±0.01 $\mu$m. Comprehensively characterizing the material in terms of crystallinity, composition and chemical state before and after the size reduction step, it can be demonstrated with some certainty that the processing to this point has no effect of any significance aside from platelet fracturing (Supplementary Figures 3 - 5). The only effect is morphological, with the material retaining crystalline and compositional homogeneity across the platelets and, importantly, at newly exposed edge sites. This can rule out the possibility of preferential breaking directions or effects like dissolution of components upon fragmentation, which would likely bring about some shift in the local crystalline structure. Additionally, an introduction of surface defects, which could effect the activity across exposed basal planes, is undetectable using Ni $2p_{3/2}$ XPS surveys.

Synthetic approaches to platelet size control have been previously demonstrated for LDH compounds. For example, Xu et al[34], demonstrated synthetic size control of MgAl LDH by subtly changing parameters such as hydrothermal treatment time and temperature during synthesis. However, many chemical properties are non-transferable when one switches between LDH compounds with different metal-centre combinations. This is the case here as similar synthetic size control techniques could not be replicated for the $Ni^{2+}$ and $Fe^{3+}$ combination. The non-amphoteric nature of $Fe^{3+}$ intermediates in the coprecipitation reaction scheme is likely the critical factor[33, 35, 36]. Controlled average particle size for colloidal dispersions can be achieved using centrifugation to produce a range of unique particle dispersions whose mean flake length ($<L>$) varies in a sequential manner (Supplementary Figure 6).[37] The process involves carefully choosing centrifugation conditions (time and rate in rotations per minute (rpm)) which will apply enough centrifugal force to the colloids to cause sedimentation of some, but not all of the dispersion. What is left after a successful centrifugation step is a dispersed 'supernatant' (smaller $<L>$) and sediments (larger $<L>$). In this work a combination of fragmentation and centrifugation steps were used to produce three particle dispersions each with a unique size distribution (Figure 3a - 3f) prepared from a range of 500 - 7000 rpm. The upper centrifugation limit was set at 7000 rpm for this work to ensure contamination from foreign particles did not become a factor when attempting to achieve accurate mass loading. Within this range it is possible to attain unique size distributions (i.e. average lateral particle size $<L>$), albeit with some overlap of standard deviations ($\sigma$). Nevertheless, polarization curves (Figure 3j) indicate a significant overpotential reduction (at 10 mA·cm$^{-2}$) for particles centrifuged at 7000 rpm. It is not always possible to observe an obvious trend in the quoted overpotentials, so it is useful to cycle the LDH electrodes using cyclic voltammetry (CV) within a fixed potential range (3g-3i) which should allow the materials to stabilize to some extent[15, 38]. This should provide a more realistic set of overpotential values for comparing electrode performances (Figure 3k). For each sample, the redox activity just below the onset overpotential for catalysis is consistent with the behaviour of the as-synthesized NiFe LDH hexagons deposited on Ni foam (Supplementary Figure 2c) and consist of $Ni^{2+}/Ni^{3+}$ redox couples in different compositional and structural environments, with subtle phase transitions being induced during early cycling to cause peak shift and splitting. Additionally, gradual growth of surface hydrous oxides on the catalyst with cycling gives the appearance of current enhancement for the respective peaks (see Supplementary Note for full details). Vitally, the effects are consistent across size selected samples and as such cannot be judged to influence the calatytic improvement for smaller platelets. In addition, there were no changes of significance in the AFM surface profiles or zeta potentials for the size-selected material (Supplementary Figure 7), meaning there can be no serious influence of surface charge or surface roughness on the relative performances. Thus, the enhancement of respective size-selected platelets is judged as an affect of their relative density of active edge sites. During early cycling one can see the overpotential fluctuating quite rapidly, reaching overpotentials more than 50 mV below initial discharge values. In this time, it is thought that a combination of subtle crystallographic phase changes and active-site degradation produce the unique shape. Phase changes have been observed for nickel-hydroxide-based OER catalysts in alkaline media. The transitions involve an initial hydrated $\alpha$-hydroxide phase which can contract to a $\gamma$-oxyhydroxide phase upon $Ni^{2+}$ oxidation (along with deprotonation of interlamellar $H_2O$), and further to a significantly denser, dehydrated $\beta$-hydroxide/oxyhdroxide crystal, as described by Bode et al.[39] More recently, Dionigi et al.[28] have demonstrated analogous phase changes within NiFe mixed-metal catalysis. It is in fact possible to detect contractions in the basal plane after cycling by way of post-mortem XRD analysis (Supplementary Figure 8), which align precisely with the expected dimensions of the $\gamma$-hydroxide phase, based on the analogy with the Bode cycle. This is the extent of phase transitions for NiFe LDH catalysis, as Trotochaud et al.[38] demonstrate for $Ni(OH)_2$, with any significant Fe inclusion the crystal does not produce a structural $\beta$-phase. The $\alpha \rightarrow \gamma$ transition generally takes place during early cycling and allows for some improvement in electrochemical performance (i.e. lower overpotentials in this case). Although the principle degradation effects causing the subsequent sharp rise in $\eta$ from cycle number four onward is not exactly known, it is likely a result of leaching of some metal centres from the catalyst crystal structure in response to OER activity over time. Beyond this, the $\eta$ values approach equilibrium for the three samples and one can start to see an apparent correlation between the overpotentials and the final centrifugation rate (and hence $<L>$). This is a subtle difference between samples but provides a proof-of-concept representation of the proposed theory, which is also reflected in cell stability tests over extended time periods (Supplementary Figure 9b). Electrochemical surface area (ECSA) analysis complements the observation as it shows no detectable increase in active surface area for size-reduced material (Supplementary Figure 10). This points towards active edge sites rather than basal plane activity as the principal area. Additionally, the process is highly reproducible over different particle size distributions. Supplementary Figure 11 represents an example of the same principle size-activity relationship but over a broader range of $<L>$. With values as low as 245±7 mV at 10 mA·cm⁻² current density (before iR-correction), the sr-NiFe LDH material prepared here represents significant improvement on reported pure materials in the literature (Table 1). For example, Jaskaniec et al[17] report 340 mV at 10 mA·cm⁻² while Diaz-Morales et al[40] report 290 mV at just 1 mA·cm⁻². In fact, the majority of literature values are reported around 300 mV for pure NiFe LDH.[9, 24] Efforts now must be made to retain the impressive catalysis of the sr-NiFe beyond its early cycle-life. The performance of composites of this nature is heavily influenced not only by the intrinsic properties of the components but also by their interfacial connectivity, and hence the method of preparation. In this sense parameters like overpotential and Tafel slope (Table 1) can give a good indication of a catalyst system's worth, but it is important to remember that there are a number of additional factors such as current density, mass loading and choice of substrate and electrolyte, which can also influence the results[41, 42]. Thus, in the following section, the effect of carbon nanotube matrices on the extended activity of sr-NiFe LDH platelets will be presented.

## 2.3  NiFe LDH – SWCNT nanocomposite

In the interest of improving cyclability and combating the $\eta$ losses, composite studies are carried out in addition to size treatment, the aim being to minimize overlap of adjacent NiFe LDH platelets while providing an integrated conductive matrix which can facilitate the rapid electrochemical mechanism. A number of interesting composite materials have been suggested and researched for LDH-based OER catalysis in recent times including various carbon allotropes[32, 43], metal-organic frameworks (MOF)[29, 44] and more novel materials, often various nanowires[45, 46]. However, CNTs have most notably been demonstrated as useful additives to this end, yielding improved overpotential, Tafel slope, conductivity and stability when coupled with NiFe LDH in alkaline electrolyzer systems. CNTs represent a well-established option as a conductive additive in electrochemistry while also being readily available. Composites of the sr-NiFe LDH and SWCNT materials were made by mixing IPA dispersions of each. In order to attain a well-integrated composite system, good connectivity needed to be achieved. However, the affinity between LDHs and SWCNT proved to be weak. Some relatively low-power tip-sonication after mixing can help achieve higher connectivity within the LDH-SWCNT matrix which is reflected electrochemically (Supplementary Figure 13). Using this approach, composites of different SWCNT weight percentages were prepared and compared by cycling 50 times under OER conditions.

According to Figure 4a it can be noted then that SWCNT additives may act to minimize inter-particle contact of sr-NiFe LDH particles within the OER electrode as well as providing the expected conductive matrix. Not surprisingly, this system leads to improved retention of the overpotential values beyond 4 – 5 cycles (Figure 4b). While trace amounts of Ni and Fe are detectable within the SWCNTs (Supplementary Figure 14), control experiments involving cycling pure SWCNTs on Ni foam indicate that no significant contribution is made to the OER (Supplementary Figure 15). Although mass loading was previously optimized as 0.12 mg·cm$^{-2}$ for pure NiFe LDH electrodes, replacing some percentage of the mass with a foreign species will likely affect this. Mass loading in the range 0.12 – 0.18 mg·cm$^{-2}$ was tested for samples containing 10, 15 and 20 weight % of SWCNT with NiFe LDH. If the mass loading is kept consistent across samples of various SWCNT wt% an inverse relationship is observed, with the electrocatalytic output decreasing (higher $\eta$ values) with more SWCNT (Figure 4b). This is expected as there would be fewer active components in each case. This can be compensated by increasing the mass loading to a particular value for each composite. The ideal loading in each case (Figure 4c) demonstrated similar electrocatalytic behaviour. Mass loading was optimised by trial.

Combining size controlled platelets in composite systems, fully optimized electrodes were prepared by mixing sr-LDH (0.16 ± 0.01 $\mu$m, 5000 rpm) in IPA with 20 wt% SWCNT, followed by composite spraying on nickel foam to 0.17 mg·cm$^{-2}$ mass loading. The final product gives an impressive electrocatalytic output with an overpotential reading of 237±7 mV vs RHE which, within the calculated uncertainty for $\eta$ values, shows little improvement on the pure size-selected sr-NiFe LDH particles. The real improvement in the composite becomes clear with cycling, as the retention through 50 cycles is the highest yet, with just 7% increase in $\eta$ (Figure 5a), a considerable long-term improvement on what could be demonstrated with the treated LDH material alone.

Upon further cycling, a stable value is reached at around 275 mV (Supplementary Figure 16). By studying Tafel slopes (Figure 5b) one can gain some insight into the improved electrocatalysis of the optimized electrode. Graphs of this kind are based on the Tafel equation (Eq.1) which relates the overpotential $\eta$ to the current density $j$ (and exchange current density $j_o$) so that the proportionality constant $b$ (Tafel slope) will give information regarding the sensitivity of the reaction rate to a change in potential across the cell:

$$\eta = b \times \log_{10} \frac{j}{j_0} \qquad (1)$$

The increase in $b$ upon initial introduction of carbon additives suggests the sensitivity is inhibited but once particle size-selection is applied, the value returns almost to that of the pure sr-NiFe. With considerable particle size reduction within a given electrode, an increase in contact area between the active material and conductive SWCNT matrix is achieved which is likely allowing for this positive affect on the Tafel slope.

This in turn suggests improved kinetics within the optimized electrode which will have a positive effect in the electrochemistry. Chronopotentiometry (Figure 5c) represents a more practical test of performance for an electrolyzer cell with no cycling, but rather an overpotential gradient when subjected to constant current density. After 1h the overpotential value remains below 280 mV with a relatively shallow slope, and remains below 295 mV after 12h (Supplementary Figure 17). This performance suggests improvement over similar NiFe LDH- carbon-based composite materials in the field of OER catalysis[5, 25, 31]. Further stability testing on timescales of one week (Supplementary Figure 18) demonstrate continued activity with overpotentials remaining below 360 mV, despite some apparent fluctuations caused by large bubbles forming on the surface of the active catalyst in the presented static setup. It is thought that at these extended timescales, the catalytic stability is governed by a number of possible factors, including continued leaching of the metal components from the structure and the growth of a surface hydrous oxide layer. In conclusion, combined post-synthesis processing techniques of size-selection and composite preparation have been combined to improve the electrocatalytic outlook for NiFe LDH material as a highly active non-precious metal-based anode in an alkaline electrolyzer cell. This presents a low-cost, scalable and low-energy process from chemical synthesis to device preparation. Electrodes were tested with the aim of determining their practical viability in operating conditions by predominantly studying cycle life rather than conventional polarization curves to understand the processes which dominate the catalytic output in the early cycle life of an electrolyzer cell. Additionally, this work suggests, from an electrochemical perspective, the existence of a majority of catalytically active sites around platelet edges with a minority along the planar surface based on the effect observed of decreasing the mean platelet sizes $<L>$ on overpotentials achieved. Overall, this study aimed to utilize this effect to offer a simple yet effective set of steps which allow notable improvements in catalytic behaviour of electrochemically active platelet-like materials which is not limited to the combination of metals considered here.

    In conclusion to the presented work, highly crystalline NiFe LDH planar hexagons with mean lateral dimensions 0.78±0.2 $\mu$m were synthesised by homogeneous precipitation and studied based on their OER electrocatalytic properties and the potential for enhancement by particle size control. Post-synthetic treatment of the material involved centrifugation-driven size selection and particle breaking using tip-sonication. The treatment methods were shown to be successful by various characterization techniques and improved the OER overpotential relative to the as-produced material. The extent to which centrifugation could be used as a size selection tool for platelets in the range 0.2 – 1.2 $\mu$m was encouraging. It was possible to selectively isolate platelet dispersions with mean flake size as small as $<L>$ = 0.15±0.01 $\mu$m by centriguation with superior overpotential compared to the bulk NiFe LDH. As well as that, size-selected samples demonstrate a

direct relationship which exists between size and catalytic behaviour. This assumption carries through in the electrochemical data at very least, with the broken particles (from tip sonication combined with centrifugal size selection) providing the best OER overpotential within the framework of this project, $\eta$ = 245±7 mV ($<L>$ = 0.2±0.01 $\mu$m). The intrinsic redox behaviour and structural phase transitions which occur within the LDH crystals during the extended catalyst lifetime are tracked and accounted for across all platelet sizes and cannot be deemed to influence the respective performance. There is still room for progress in terms of preparing electrochemical cells which can emulate the most competitive overpotentials reported in literature to date. For instance, catalysts based on NiFe LDH composites were studied as well for their catalytic enhancement properties. Composites based on carbon nanotubes increased the electron transport capability of the material. The results suggest that a well connected system can be achieved between the active component and carbon additive without the necessity of a bottom-up approach involving direct growth onto a carbon network. In fact, sr-NiFe LDH appears quite compatible in composite systems with SWCNTs in weight ratios of 10% - 20%. Composites prepared as such for this work ($<L>$ = 0.16±0.005 $\mu$m, 20 wt% SWCNT) displayed excellent cycling performance compared with the active material alone, capable of lower stable overpotentials beyond 150 cycles. Further OER enhancement may result from achieving complete homogeneous deposition of NiFe LDH platelets onto Ni foam substrates with mass loading between 0.1 and 0.2 mg.cm$^{-2}$ and without agglomeration. This was achieved by using ultrasonic tip spraying. This should allow more complete deposition of the active material across the entire electrode surface and so produce more competitive overpotential values in relation to ongoing research in the field of OER catalysis.

## Methods

**NiFe LDH Synthesis** Ni(NO$_3$)$_2$·6H$_2$O, Fe(NO$_3$)$_3$·9H$_2$O, urea and triethanolamine (TEA) were made into a 200 ml solution using deionized (DI) water such that their respective concentrations were 7.5 mM, 2.5 mM, 17.5 mM and 10 mM, respectively. The resulting solutions were stirred at 150 rotations per minute (rpm) at room temperature for 24 h, after which a brown precipitate was observed. Next, the reaction mixture was split into two 100 mL round bottom flasks (80 mL in each) and heated under reflux to 100°C in an oil bath for 48h. Afterwards the system was allowed to cool naturally to room temperature. The obtained dispersion was then centrifuged at 3000 rpm for 10 minutes to precipitate the NiFe LDH platelets. The prepared material was then washed three times by dispersing the sediments in DI water by shaking, followed by centrifugation (5000 rpm for ten minutes using a Heraeus Multifuge XI Centrifuge). The clean powders were then washed similarly in IPA before re-dispersing and storing for further use. All chemicals were purchased from Sigma-Aldrich and used without any modification.

**NiFe LDH Characterization** The clean material powders dispersed in IPA were then characterized by means of SEM, TEM, XRD and SAED to determine chemical composition, crystallinity and morphology. SEM was carried out using a Zeiss Ultra Plus field-emission microscope with a Gemini column and a secondary-electron detector, operating in accelerating voltage range 2–5 keV. TEM images were taken using an FEI Titan 80 – 300 kV FEG STEM microscope with SAED acquired in-situ in this machine. Powder XRD was carried out using a Bruker Advance Powder X-ray diffractometer with a molybdenum K-$\alpha$ emission source in the Bragg-Brentano configuration. XPS measurements were performed using an Omicron XPS model EA125 with a monochromatic Al K$\alpha$ source, where the binding energy was calibrated to the C 1s peak.

**Centrifugal Size-Selection** Sample dispersion of NiFe LDH in IPA were size-selected by initially centrifuging at 500 rpm using a Heraeus Multifuge X1 Centrifuge with a F15-6x100y carbon fibre rotor.

**Electrodes Manufacturing** Electrodes for OER testing were prepared by spraying NiFe LDH component onto a nickel foam using a USI Prism Ultracoat 300 spray tool with the substrate kept at 100°C. A flow rate of 0.5 mL/min was used and mass loading was monitored using a Sartorius SE2 ultra-microbalance.

**SWCNT Dispersions Preparation** Dispersions of SWCNTs in IPA were prepared by adding 10 mg of SWCNT powder (P3-SWCNT, Carbon Solutions Inc. - Supplementary Figure 14) to 100 ml of IPA and applying ultrasonication by way of tip sonication (Fischer Scientific Sonic Dismembrator Ultrasonic Processor) at 40% power for 45 min assisted by a cooling system maintaining the dispersion at 5°C. Following that, bath sonication at 37 kHz was applied to the dispersion for 1 hour using a Fisherbrand 112xx Series Advanced Ultrasonic Cleaner at 60% power.

**Composite Preparation** Composites were prepared using known concentrations of SWCNT and NiFe LDH suspensions in isopropanol (IPA). Appropriate volumes of each were measured out using glass pipettes in order to achieve 10 mL samples of 10, 15 and 20 weight% SWCNT wrt NiFe LDH. In order to achieve good contact between the components, further tip sonication at 40% power for 10 minutes was applied.

**Electrochemical Testing** Electrodes were placed in a three-electrode setup with a platinum wire acting as the counter and Ag/AgCl reference electrode (3.5 M KCl filling solution). The electrolyte used was 1 M potassium hydroxide (KOH) solution. Cyclic voltammetry, linear voltammetry and chronopotentiometry were utilized as the principle characterization tools in terms of electrochemistry. The potentiostat used was a BioLogic VMP 300. Voltage sweep rates were kept at 5 mV·s$^{-1}$ in a potential window 0 – 0.6 V (vs Ag/AgCl) for all voltamograms. The potential measured vs Ag/AgCl is mathematically converted to potential vs reversible hydrogen electrode (RHE) according to Eq 2:

$$E_{\text{RHE}} = E_{\text{Ag/AgCl}} + 0.059 \cdot \text{pH} + E^{\text{o}}_{\text{Ag/AgCl}} \tag{2}$$

where $E^{o}_{Ag/AgCl}$ is the saturated Ag/AgCl electrode potential equal to 0.197 V at 25 °C.

## Data Availability

The data relating to the findings of this work is available from the corresponding author, subject to reasonable request.

## Acknowledgements

D.T., S.J., J.C. and V.N. wish to thank the support of the ERC CoG, 3D2DPring (GA 681544) and PoC Powering eTextiles (GA 861673) and SFI AMBER (12/RC/2278 P2). M.P.B acknowledges the financial support of the European Union's Horizon 2020 Research and innovation programme under the Marie Sk-lodowska-Curie Actions IF under the project TriCat4Energy (Grant Agreement no. 884318). The authors would like to thank the Advanced Microscopy Lab and CRANN Trinity College Dublin for providing STEM-EDX measurements.

This publication has emanated from research supported in part by a grant from Science Foundation Ireland under Grant number 12/RC/2278 P2. For the purpose of Open Access, the author has applied a CC BY public copyright licence to any Author Accepted Manuscript version arising from this submission.

## Competing Interests

There are no financial or non-financial conflicts to declare.

## Author Contributions

D.T., J.C., M.P.B., S.J. and V.N. took part in discussion and proposal of project structure and planning experiments. Synthesis of NiFe LDH was carried out by D.T. and S.J. SEM was performed by D.T, as well as subsequent analysis of platelet sizes. Electrochemistry was performed by D.T. with assistance from J.C. and M.P.B. TEM analysis was carried out by A.R. along with in-situ EDX mapping and SAED. Diffraction patterns acquired by XRD were performed by B.S. and D.T. XPS surveys were acquired by L.G. and C.M., and fitted by M.P.B. AFM analysis was performed by K.O. and G.S.D. Zeta potentials and TGA measurements were performed by D.T and S.J. Electrochemical data interpretation was carried out by J.C., M.P.B. and D.T. and wider outlook and conclusions were discussed by J.C., M.P.B., S.J., V.N. and D.T. The paper was written by D.T. with assistance from J.C., and all authors contributed to the manuscript.

Table 1: Comparative OER activities for various competitive NiFe LDH catalyst systems (* this work).

| Catalyst | η @ 10 mA.cm$^{-2}$ (mV vs RHE) | Current density (mA.cm$^{-2}$) | Tafel slope (mV.dec$^{-1}$) | Mass loading (mg.cm$^{-2}$) | Substrate |
|---|---|---|---|---|---|
| IrO$_2$[24] | 338 | 10 | 47 | 0.21 | GC |
| NiFe LDH[17] | 360 | 5 | n/a | 0.12 | NF |
| NiFe Nanosheets[24] | 302 | 9.4 | 40 | 0.07 | GC |
| NiFe-CNT[25] | 220 | 10 | 31 | 0.25 | GC |
| NiFe-rGO[31] | 230 | 10 | 42 | 0.25 | GC |
| NiFe@NiCoP[5] | 220 | 10 | 49 | 2 | NF |
| **sr-NiFe**[*] | 245 | 10 | 32 | 0.16 | NF |
| **sr-NiFe-CNT**[*] | 237 | 10 | 32 | 0.16 | NF |

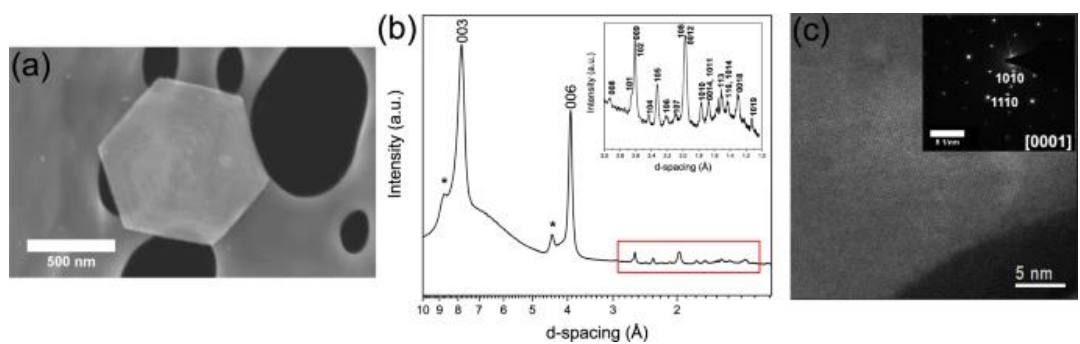

Figure 1: Structural characterization of NiFe LDH hexagonal platelets; (a) SEM micrograph, (b) Mo K$\alpha$-source XRD diffraction pattern (inset, zoomed signals for the 1.3-3 Å d-spacing region), (c) atomic resolution HAADF-STEM (inset, SAED pattern) for as-synthesized NiFe LDH. Scale bars are (a) 500 nm, (c) 5 nm and inset, 5 1/nm.

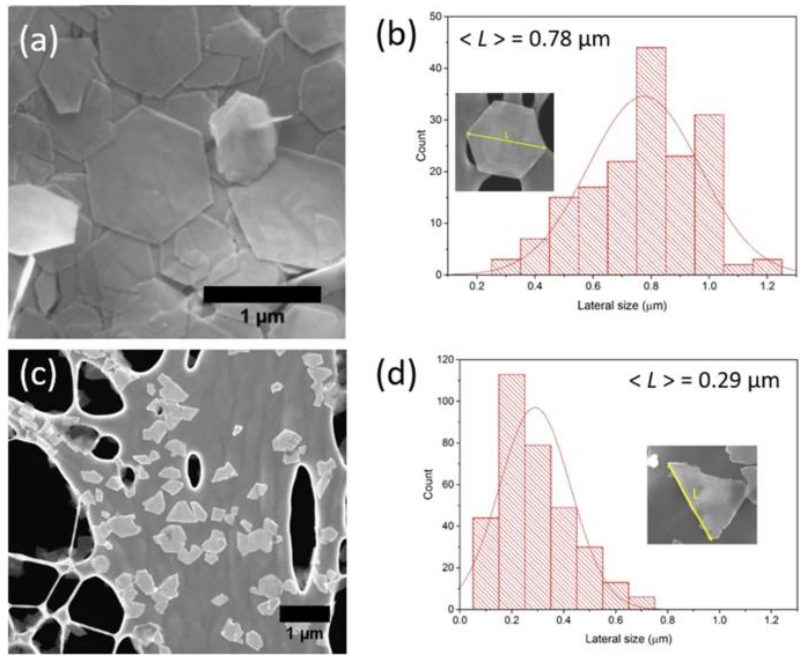

Figure 2: Demonstration of platelet size reduction upon sonication treatment; (a) and (c) SEM micrographs of as-produced and size reduced platelets respectively and (b) and (d) associated platelet size distributions indicating mean platelet size. (Inset (b) and (d) indicate the method of measuring lateral platelets size L). Scale bars are 1 $\mu$m in (a) and (c).

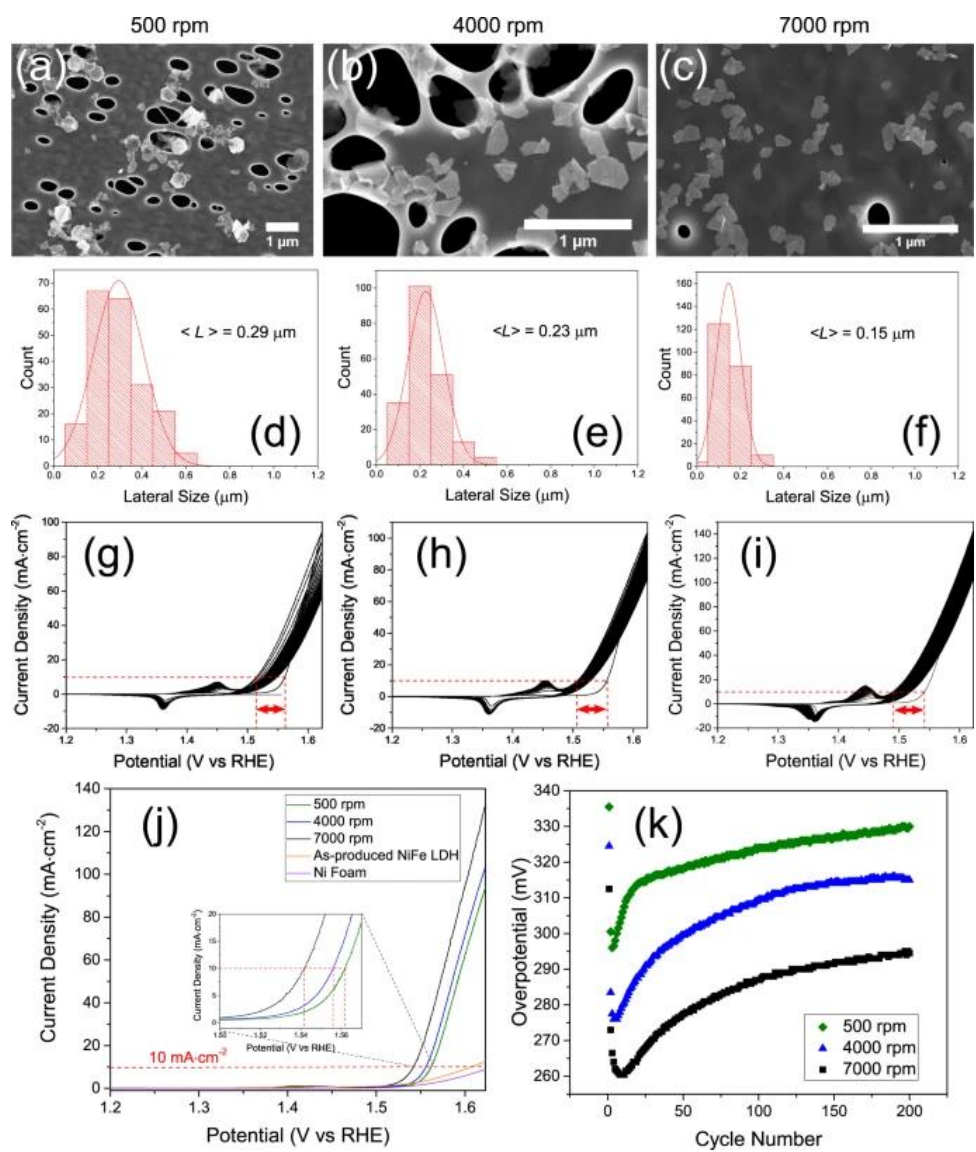

Figure 3: Summary of the relationship between controlled platelet sizes and electrocatalytic output; (a) - (c) SEM micrographs and (d) - (f) accompanying size-distribution histograms for the sr-NiFe LDH with final centrifugation rates 500, 4000 and 7000 rpm respectively. (g) – (i) CVs over 200 cycles for each of the respective electrodes. Quoting the anodic potentials at 10 mA.cm$^{-2}$ of successive cycles allowed the overpotentials to be tracked during an extended catalyst lifetime. (j) Initial polarization curves for electrodes prepared with each dispersion and (k) overpotentials as a function of cycle number, averaged over two electrode samples in each case (full data in Supplementary Figure 9a). Scale bars are 1 $\mu$m in (a), (b) and (c).

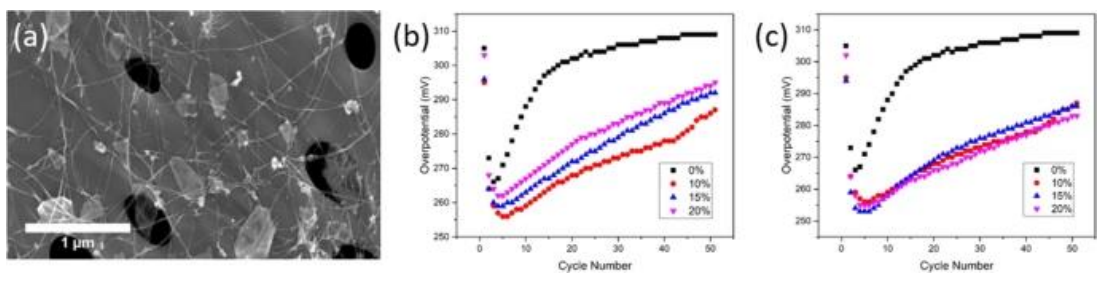

Figure 4: Optimizaton of integrated composite materials; (a) SEM of sr-NiFe-SWCNT composite on lacey carbon and (b) comparison of overpotentials through 50 cycles of OER electrodes prepared using sr-NiFe with 10, 15 and 20 weight % SWCNT with identical mass loading and (c) optimized mass loading. Scale bar is 1 $\mu$m in (a).

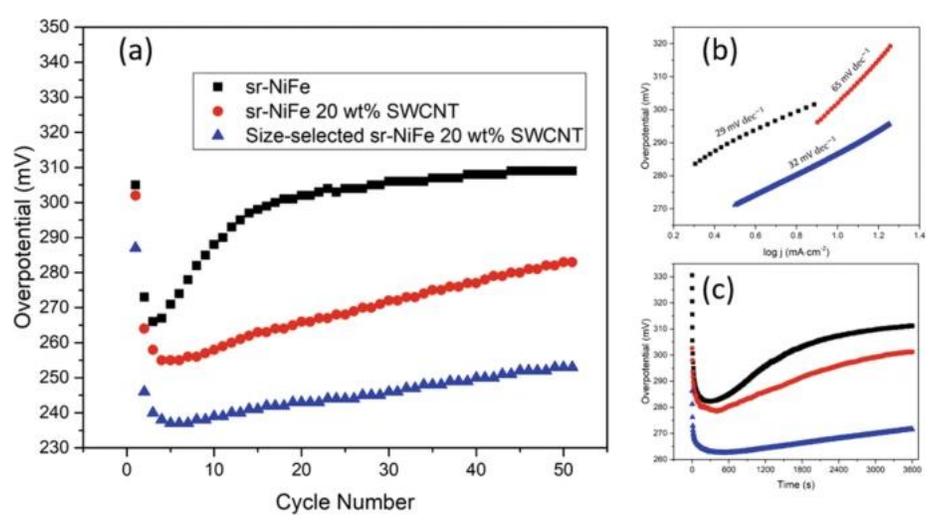

Figure 5: Performance summary for 'fully-optimized' catalysts in the scope of this work. Comparison of (a) overpotentials through 50 cycles and (b) Tafel slopes of OER electrodes prepared using size-selected sr-NiFe LDH with 20 wt% SWCNT, a non-size-selected composite and pure sr-NiFe LDH. (c) Chronopotentiometry curve of the optimized electrode through 1 hour at 10 mA·cm$^{-2}$.